\documentclass[11pt, epsfig]{article}
\usepackage{color}

\usepackage{graphicx}
  \usepackage{amsthm,amsfonts}
  \usepackage{amsmath}
\newcommand{\bea}   {\begin{eqnarray}}
\newcommand{\eea}   {\end{eqnarray}}

\begin{document}
\renewcommand{\thefootnote}{\fnsymbol{footnote}}

\thispagestyle{empty}

\title{Invariant PDEs of Conformal Galilei Algebra as deformations: cryptohermiticity and contractions}

\author{N. Aizawa\thanks{{E-mail: {\em aizawa@mi.s.osakafu-u.ac.jp}}},\quad 
Z. Kuznetsova\thanks{{E-mail: {\em zhanna.kuznetsova@ufabc.edu.br}}}
\quad and\quad F.
Toppan\thanks{{E-mail: {\em toppan@cbpf.br}}}
\\
\\
}
\maketitle

\centerline{$^{\ast}$
{\it Department of Mathematics and Information Sciences,}}
{\centerline {\it\quad
Graduate School of Science, Osaka Prefecture University, Nakamozu Campus,}}
{\centerline{\it\quad Sakai, Osaka 599-8531 Japan.}}
\centerline{$^{\dag}$
{\it UFABC, Av. dos Estados 5001, Bangu,}}{\centerline {\it\quad
cep 09210-580, Santo Andr\'e (SP), Brazil.}
\centerline{$^{\ddag}$
{\it CBPF, Rua Dr. Xavier Sigaud 150, Urca,}}{\centerline {\it\quad
cep 22290-180, Rio de Janeiro (RJ), Brazil.}
~\\
\maketitle
\begin{abstract}

We investigate the general class of second-order PDEs, invariant under the $d=1$ $\ell=\frac{1}{2}+{\mathbb N}_0$ centrally extended Conformal Galilei Algebras, pointing out that they are deformations of decoupled systems. For $\ell=\frac{3}{2}$ the unique 
deformation parameter $\gamma$ belongs to the fundamental domain $\gamma\in ]0,+\infty[$. \par
We show that, for any $\gamma\neq 0$, invariant PDEs with discrete spectrum (either bounded or unbounded) induce cryptohermitian operators possessing the same spectrum as two decoupled oscillators, provided that their frequencies are in the special ratio $r=\frac{\omega_2}{\omega_1}=\pm\frac{1}{3},\pm 3$ (the negative energy solutions correspond to a special case of Pais-Uhlenbeck oscillator), where
$\omega_1,\omega_2$ are two different parameters of the invariant PDEs.\par
We also consider the $\gamma=0$ decoupled system for any value $r$ of the ratio. It possesses enhanced symmetry at the critical values $r=\pm \frac{1}{3}, \pm 1,\pm 3$. Two inequivalent $12$-generator symmetry algebras are found at $r =\pm\frac{1}{3},\pm 3$ and $r=\pm 1$, respectively. The $\ell=\frac{3}{2}$ Conformal Galilei Algebra is not a subalgebra of the decoupled symmetry algebra. Its $\gamma\rightarrow 0$ contraction corresponds to a $8$-generator subalgebra of the decoupled $r=\pm\frac{1}{3},\pm 3$ symmetry algebra.\par
The features of the $\ell\geq \frac{5}{2}$ invariant PDEs are briefly discussed.
~\\\end{abstract}
\vfill

\rightline{CBPF-NF-003/15
}

\newpage
\section{Introduction}

In \cite{Aizawa:2013vma} and \cite{Aizawa:2014uma} second-order PDEs, invariant under the ${\widehat{\mathfrak{cga}}}_\ell$ ($\ell=\textstyle{\frac{1}{2}}+\mathbb{N}_0$) centrally extended Conformal Galilei Algebra \cite{negro1997nonrelativistic}, were constructed. They were shown to possess a spectrum which is either continuous \cite{Aizawa:2013vma} or discrete (positive and bounded) \cite{Aizawa:2014uma}.  
In \cite{Aizawa:2013vma} the invariant PDEs were obtained via Verma module representation, while in \cite{Aizawa:2014uma}  the so called on-shell condition was used (for the cases at hand the two approaches are proven to be equivalent). \par
In this paper we address several important issues that were not touched in these  two previous works. We name a few: the identification of the general class of invariant PDEs (which turns out to depend on real parameters  belonging  to a fundamental domain),  the existence of a contraction algebra, the reason for the cryptohermiticity (we use here the word adopted in \cite{smilga2009comments}) of the discrete spectrum, the construction of the {{Hilbert space connected with Pais-Uhlenbeck oscillators with unbounded spectrum, etc.}
\par
Specifically, the following list of results is derived in the present paper (we limit here in the Introduction to discuss the first non-trivial case obtained for ${\ell}={\textstyle{\frac{3}{2}}}$, the ${\ell}>{\textstyle{\frac{3}{2}}}$ cases are commented in Section {\bf 8}): two special differential realizations of ${\widehat{\mathfrak{cga}}}_\ell$ produce, as invariant PDEs,
Schr\"odinger equations with continuous and respectively discrete spectrum. Both realizations depend on a parameter $\gamma\neq0$. Unitarily inequivalent theories are recovered for $\gamma$ belonging to the fundamental domain $\gamma\in ]0,+\infty[$.\par
The $\gamma=0$ PDEs are decoupled equations. The continuum spectrum case corresponds to the free Schr\"odinger equation in $1+1$ dimensions, while the discrete spectrum case corresponds to a system of two decoupled {\em cryptohermitian oscillators} (namely, despite being non hermitian, possessing the same spectrum as two decoupled oscillators with the given frequencies). The parameter $\gamma$ can therefore be regarded as a deformation parameter and as a coupling constant.\par
Without loss of generality we can fix $\omega_1=1$ to be the energy mode of the first oscillator in the coupled PDE. Then, the ${\widehat{\mathfrak{cga}}}_{\textstyle{\frac{3}{2}}}$ invariance of the PDE is recovered if  the energy mode of the second oscillator possesses the critical values
$\omega_2=\pm {\frac{1}{3}}, \pm 3$ ($\omega_2=3$ is the solution given in \cite{Aizawa:2014uma}). The negative values correspond to an unbounded spectrum and, as explained later, are connected with special cases of the Pais-Uhlenbeck oscillators. At fixed $\omega_{1,2}$, the spectrum of the operators does not depend on the value of $\gamma$.\par
The $\gamma\rightarrow 0$ limit of the ${\widehat{\mathfrak{cga}}}_{\textstyle{\frac{3}{2}}}$ algebra produces a contraction algebra which is a symmetry subalgebra of the decoupled systems. For the decoupled oscillators (without loss of generality the analysis can be limited to the $\omega_1=1$, $\omega_2\geq 1$ domain), the PDE possesses a $9$-generator symmetry algebra at generic values, with enhanced symmetry at the critical values $\omega_2=1$ and $\omega_2=3$ (two different $12$-generator symmetry algebras are obtained at these special points). The $\gamma\rightarrow 0$ contraction algebra is a $8$-generator subalgebra of the $\omega_2=3$, $12$-generator decoupled symmetry.\par
%For all critical values $\omega_2=\pm{\frac{1}{3}},\pm{3}$ and for all values of $\gamma$ (including %$\gamma=0$), the cryptohermitian operators associated with the discrete spectrum act on the Hilbert space %${\mathcal L}^2({\mathbb R}^2)$. The existence of similarity transformations prove that the spectrum is %independent of $\gamma$. Unitary transformations change the phase of $\gamma$. Therefore, %inequivalent cryptohermitian operators with the same spectrum are labeled by $\gamma\in [0,+\infty[$.%\par
%It is easily shown that the eigenvectors are not normalized in ${\mathcal L}^2({\mathbb R}^2)$. A different %Hilbert space, ${\mathcal L}^2({\mathbb {\widetilde R}}^2)$, can be introduced. It is defined by preserving %the canonical commutation relations while changing the conjugation properties of the creation/annihilation %operators. Since the canonical commutation relations are unchanged, the spectrum of the operators acting %on
%${\mathcal L}^2({\mathbb {\widetilde R}}^2)$ coincides with the spectrum of the operators acting on
%${\mathcal L}^2({\mathbb R}^2)$. In ${\mathcal L}^2({\mathbb {\widetilde R}}^2)$ the $\gamma=0$ %operator is hermitian and given by the sum of two decoupled harmonic oscillators. The $\gamma\neq 0$ %operators are cryptohermitian and their eigenstates, which belong to ${\mathcal L}^2({\mathbb {\widetilde %R}}^2)$, are not orthogonal. 
%\par
In this paper we discuss the subtle connection with Pais-Uhlenbeck oscillators.
The Pais-Uhlenbeck model is a higher derivative system \cite{pais1950field,smilga2009comments,ketovMichiaki} which admits, via the Ostrogradski\u{\i} construction \cite{ostrogradski1850member},
a Hamiltonian formulation. The Ostrogradski\u{\i} Hamiltonian is canonically equivalent to a set of decoupled harmonic oscillators with alternating (positive and negative) energy modes. 
In a series of papers \cite{galajinsky2013dynamical2,andrzejewski2014conformal,andrzejewski2014conformal2,andrzejewski2014hamiltonian,galajinsky2015dynamical} 
the Pais-Uhlenbeck oscillators with energy modes given (up to a normalization factor) by the arithmetic progression $\omega_i = 2i-1$ were linked to the Conformal Galilei Algebras ${\widehat{\mathfrak{cga}}}_\ell$ (with $\ell = n-\textstyle{\frac{1}{2}}$).
As discussed in Section {\bf 8} the connection is rather subtle. To avoid any confusion, we should stress that the symmetry operators considered here are required to be realized as {\em first-order} differential operators. \par
%The PDE, invariant under the Conformal Galilei Algebra, is obtained for the coupled cryptohermitian operator %with $\gamma\neq 0$ and unbounded spectrum.  The derivation of the Pais-Uhlenbeck oscillator requires %two non-trivial passages which (both) spoil the Conformal Galilei invariance: {\em i}) taking the %$\gamma\rightarrow 0$ decoupling limit and {\em ii})  change the conjugation properties, by replacing the %decoupled cryptohermitian operator with the hermitian decoupled harmonic oscillator.\par
%Another result presented in the paper is the realization of a commutative diagram relating, via similarity %transformations and a change of the time coordinate, the differential realizations for coupled and decoupled %Schr\"odinger equations with continuous and discrete spectrum.\par
The scheme of the paper is as follows: in Section {\bf 2} we present the ($\gamma\neq 0$-dependent) differential realization for the deformation of the free Schr\"odinger equation at ${\ell}={\textstyle\frac{3}{2}}$ and the differential realization for the coupled oscillator. The connection of the two differential realizations obtained by similarity transformations and change of the time coordinate is shown in Section {\bf 3}.  In Section {\bf 4} the most general solution of the ${\widehat{\mathfrak{cga}}}_{\frac{3}{2}}$-invariant oscillator is given. The symmetry of the decoupled oscillator (with enhanced critical points at $\omega_2=1 $ and $\omega_2=3$)
is presented in Section {\bf 5}. The ${\ell}={\textstyle{\frac{3}{2}}}$ contraction algebra in the $\gamma\rightarrow 0$ limit is given in Section {\bf 6}. {{The Hilbert space for the oscillators and the relation with PT-symmetry
is discussed in Section {\bf 7}}}. In Section {\bf 8} the extension to the ${\ell}>{\textstyle\frac{3}{2}}$ cases and the relation to Pais-Uhlenbeck oscillators are commented. Generalizations of the present construction are discussed in the Conclusions.

\section{Differential realizations of the Conformal Galilei Algebra ${\widehat{\mathfrak{cga}}}_{\frac{3}{2}}$ }
The $d=1$ ${\ell}=\frac{3}{2}$ centrally extended Conformal Galilei algebra
${\widehat{\mathfrak{cga}}}_{\frac{3}{2}}$ consists of eight generators ($z_0,z_\pm, w_{\pm 1}, w_{\pm 3}, c$), obeying the following
non-vanishing commutation relations 
\bea\label{cga32}
\relax [z_0,z_\pm]&=& \pm 2i  z_\pm,\nonumber\\
\relax [z_+, z_-]&=& -4i z_0,\nonumber\\
\relax[ z_0,  w_k] &=& i k {w_k},\nonumber\\
\relax [ z_\pm,  w_k]&=& i(k\mp3) w_{k\pm2},\nonumber\\
\relax [ w_{|k|}, w_{-|k|}]&=&(3-2|k|){16} { c}.
\eea
We introduce in this Section two differential realizations of the above algebra.
As it will be explained in the following, the first differential realization corresponds to a symmetry of a deformed free Schr\"odinger equation in $1+1$ dimensions, while the second differential realization corresponds to a symmetry of another system 
(possessing the same spectrum as two decoupled oscillators).\par
The first differential realization is a one-parameter ($\gamma\neq 0$) extension of the differential
realization obtained in \cite{Aizawa:2013vma} via standard left action on the space of functions defined on the coset (see also \cite{gomis2012schrodinger}). The presence of $\gamma$ can be understood by the fact that the combined rescalings of the space and time coordinates which preserve the conformal Galilei algebra structure lead to just one free parameter. In
this case we obtain first-order differential operators acting on functions of $\tau, x, y$. They are given by
\bea\label{freerep}
{\overline z}_+&=& \partial_\tau,\nonumber\\
{\overline z}_0&=&-2i\tau\partial_\tau -ix\partial_x-3iy\partial_y-2i,\nonumber\\
{\overline z}_-&=&-4\tau^2\partial_\tau-4(\tau x-\frac{3}{\gamma}y)\partial_x -12\tau y\partial_y -8(\tau-ix^2),\nonumber\\
{\overline w}_{+3}&=& \partial_y,\nonumber\\
{\overline w}_{+1}&=& -2i\tau\partial_y+\frac{2i}{\gamma}\partial_x,\nonumber\\
{\overline w}_{-1}&=& -4\tau^2 \partial_y+\frac{8}{\gamma}\tau\partial_x-\frac{8i}{\gamma}x,\nonumber\\
{\overline w}_{-3}&=& 8i\tau^3 \partial_y -\frac{24i}{\gamma}\tau^2\partial_x -48(\frac{1}{\gamma}\tau x +\frac{1}{\gamma^2}y),\nonumber\\
{\overline c}&=&\frac{1}{\gamma^2}.
\eea
The second differential realization is the $\gamma$ extension of the differential realization presented in \cite{Aizawa:2014uma}.
It is given by the first-order differential operators acting on functions of $t, x, y$,
%zzzrev
\bea\label{oscrep}
{\widehat z}_0&=&\partial_t,\nonumber\\
{\widehat z}_+ &=& e^{2i t}(\partial_t+ ix\partial_x+ 3iy\partial_y+ix^2+2i),\nonumber\\
{\widehat z}_-&=&e^{-2it}(\partial_t-ix\partial_x-3iy\partial_y+\frac{12 }{\gamma}y\partial_x+7ix^2+
\frac{12 }{\gamma} xy-2i),\nonumber\\
{\widehat w}_{+3}&=& e^{3it}\partial_y,\nonumber\\
{\widehat w}_{+1}&=& e^{it}(\partial_y+\frac{2i}{\gamma}\partial_x+\frac{2i}{\gamma}x),\nonumber\\
{\widehat w}_{-1}&=&e^{-it}(\partial_y+\frac{4i}{\gamma}\partial_x-\frac{4i}{\gamma}x),\nonumber\\
{\widehat w}_{-3}&=& e^{-3it}(\partial_y+\frac{6i}{\gamma}\partial_x-\frac{18i}{\gamma}x-\frac{48}{\gamma^2}y),\nonumber\\
{\widehat c}&=& \frac{1}{\gamma^2}.
\eea
We consider now the three elements $\Omega_0, \Omega_{\pm 1}$ of the ${\widehat{\mathfrak{cga}}}_{\frac{3}{2}}$ enveloping algebra of degree $0,\pm 1$ (measured by the degree generator $-\frac{i}{2}z_0$), respectively:
\bea\label{omegafreerep}
{\Omega}_{+1}&=&  i {z}_++\frac{\gamma^2}{16} \left(\{{w}_{+3}, { w}_{-1}\}-\{{w}_{+1}, { w}_{+1} \}\right),\nonumber\\
{\Omega}_{0}&=& 
i {\overline z}_0+\frac{\gamma^2}{32} \left(\{{w}_{+3}, { w}_{-3}\}-\{{ w}_{+1}, {w}_{-1} \}\right),\nonumber\\
{ \Omega}_{-1}&=& 
i {\overline z}_{-}+\frac{\gamma^2}{16} \left(\{{w}_{+1}, { w}_{-3}\}-\{{w}_{-1}, { w}_{-1} \}\right),
\eea
where the curly brackets on the r.h.s. denote anticommutators.\par
The three operators ${\Omega}_{\pm 1,0}$ close the ${\mathfrak{sl}(2)}$ algebra, with ${\Omega}_0$ the Cartan element:
\bea
\relax [{\Omega}_0,{\Omega}_{\pm1}]&=& \mp 2{\Omega}_{\pm 1},\nonumber\\
\relax[{\Omega}_{+1},{\Omega}_{-1}]&=& 4{\Omega}_0.
\eea
It is important to note that for both differential realizations (\ref{freerep}) and (\ref{oscrep}), $\Omega_0, \Omega_{\pm 1}$ are presented 
as second order differential operators.\par
For the differential realization (\ref{freerep}) we have
\bea\label{omegafreerep}
{\overline \Omega}_{+1}&=& i\partial_\tau -i\gamma x\partial_y + \frac{1}{2}\partial_x^2=i {\overline z}_+-\overline H_+,\nonumber\\
{\overline \Omega}_{0}&=& -2i\tau {\overline \Omega}_{+1}=i {\overline z}_0-\overline H_0,\nonumber\\
{\overline \Omega}_{-1}&=& -4\tau^2 {\overline \Omega}_{+1}=i {\overline z}_--\overline H_-.
\eea
The ${\widehat{\mathfrak{cga}}}_{\frac{3}{2}}$ on-shell invariant condition for ${\overline\Omega}_{\pm 1,0}$ (see \cite{Niederer:1972zz,Toppan:2015qla,Aizawa:2014uma} for a definition) is guaranteed by the fact that their only non-vanishing commutators with the ${\widehat{\mathfrak{cga}}}_{\frac{3}{2}}$ generators are expressed as
\bea
\relax [\overline z_0, {\overline{\Omega}}_{+1}]&=& 2i {\overline\Omega}_{+1} , \nonumber\\
\relax [\overline z_-, {\overline{\Omega}}_{+1}]&=& 4i{\overline \Omega}_0= 8\tau {\overline\Omega}_{+1} , \nonumber\\
\relax [\overline z_+, {\overline{\Omega}}_{0}]&=& -2i{\overline \Omega}_{+1}= \tau^{-1}{\overline\Omega}_0 , \nonumber\\
\relax [\overline z_-, {\overline{\Omega}}_{0}]&=& 2i{\overline \Omega}_{-1}= 4\tau{\overline\Omega}_0 , \nonumber\\
\relax [\overline z_+, {\overline{\Omega}}_{-1}]&=& -4i{\overline \Omega}_{0}=  2\tau^{-1}{\overline\Omega}_{-1}, \nonumber\\
\relax [\overline z_0, {\overline{\Omega}}_{-1}]&=& -2i{\overline \Omega}_{-1}.
\eea
The degree $1$ invariant equation
\bea\label{deforfree}
{\overline\Omega}_{1}\Psi(\tau,x,y)=0 \quad &\Rightarrow & i\partial_\tau\Psi =-\frac{1}{2}\partial_x^2\Psi 
+i\gamma x\partial_y\Psi 
\eea
is a Schr\"odinger equation with $\tau$ playing the role of a time coordinate. The parameter $\gamma$ is a coupling constant. This equation can be regarded as a $\gamma$-deformation of the free Schr\"odinger equation in $1+1$ dimensions.
\par
For the differential realization (\ref{oscrep}), $\Omega_0, \Omega_{\pm 1}$ are given by
\bea\label{omegaoscrep}
{\widehat{\Omega}}_{+1}&=& e^{2it}{\widehat \Omega}_0=i {{\widehat z}}_+- {\widehat H}_+,\nonumber\\
{ \widehat{\Omega}}_{0}&=& i\partial_t +\frac{1}{2}{\partial_x}^2-\frac{1}{2}x^2 -3y\partial_y-i\gamma x\partial_y-\frac{3}{2}=i { {\widehat z}}_0-{\widehat H}_0,\nonumber\\
{\widehat{\Omega}}_{-1}&=& e^{-2it}{\widehat \Omega}_0=i {{\widehat z}}_--{\widehat H}_-.
\eea

The on-shell invariant condition is guaranteed by the fact that their only non-vanishing commutators with the ${\widehat{\mathfrak{cga}}}_{\frac{3}{2}}$ generators are expressed as
\bea
\relax [ {\widehat z}_0, {{{\widehat \Omega}}}_{+1}]&=& 2i {{\widehat \Omega}}_{+1} , \nonumber\\
\relax [ {\widehat z}_-, {{{\widehat \Omega}}}_{+1}]&=& 4i{{\widehat \Omega}}_0=4i e^{-2it} {{\widehat \Omega}}_{+1} , \nonumber\\
\relax [{\widehat  z}_+, {{{\widehat \Omega}}}_{0}]&=& -2i{{\widehat \Omega}}_{+1}= -2ie^{2it}{{\widehat \Omega}}_0 , \nonumber\\
\relax [ {\widehat z}_-, {{ {\widehat \Omega}}}_{0}]&=& 2i{{\widehat \Omega}}_{-1}= 2i e^{-2it}{{\widehat \Omega}}_0 , \nonumber\\
\relax [ {\widehat z}_+, {{ {\widehat \Omega}}}_{-1}]&=& -4i{{\widehat \Omega}}_{0}=  -4ie^{2it}{{\widehat \Omega}}_{-1}, \nonumber\\
\relax [{\widehat z}_0, {{ {\widehat \Omega}}}_{-1}]&=& -2i{{\widehat \Omega}}_{-1}.
\eea

The degree $0$ invariant equation
\bea\label{cryptoosc}
{{\widehat \Omega}}_{0}\Psi(t,x,y)=0 \quad &\Rightarrow& i\partial_t\Psi =\left(
-\frac{1}{2}{\partial_x}^2+\frac{1}{2}x^2 +3y\partial_y+i\gamma x\partial_y+\frac{3}{2}\right)\Psi
\eea
is a Schr\"odinger equation with $t$ playing the role of the time coordinate. The non-hermitian operator in the right hand side possesses a discrete and positive spectrum. The parameter $\gamma$ is a coupling constant. The equation (\ref{cryptoosc}) can be regarded as a $\gamma$-deformation of a decoupled ``cryptohermitian oscillator" discussed in the following.

\section{Connection of the two differential realizations}
The two differential realizations (\ref{freerep}) and (\ref{oscrep}) of the ${\widehat{\mathfrak{cga}}}_{\frac{3}{2}}$ algebra introduced in Section {\bf 2}  induce Schr\"odinger equations from, respectively, 
degree $1$ and degree $0$ invariant operators.\par
The two differential realizations are connected via a similarity transformation coupled with a redefinition of the time coordinate.\par
Let us denote as ${\widehat {g}}$ an operator entering (\ref{oscrep}) or (\ref{omegaoscrep}) and as ${\overline g}$ its  corresponding operator
entering (\ref{freerep}) or (\ref{omegafreerep}). For convenience we introduce the operator ${\widehat X}_+$ by setting, for ${\widehat z}_+$ in (\ref{oscrep}),
\bea
{\widehat z}_\pm = e^{\pm 2it}(\partial_t+{\widehat X}_\pm),\quad && {\widehat X}_+= ix\partial_x+ 3iy\partial_y+ix^2+2i.
\eea

The connection is explicitly realized by the similarity transformation
\bea
{\widehat g}\mapsto {\overline g} &= &e^S {\widehat g} e^{-S},\quad \quad (e^S= e^{S_2}e^{S_1}),\nonumber\\
S_1&=& t{\widehat X}_+,\nonumber\\
S_2&=&\frac{1}{2}x^2,
\eea
 supplemented by the redefinition of the time coordinate
\bea
t\mapsto\tau&=& \frac{i}{2}e^{-2it}.
\eea
The first similarity transformation (induced by $S_1$) allows to map
\bea
{\widehat z}_+\mapsto \widetilde{z}_+ &=&e^{S_1}{\widehat z}_+e^{-S_1}= e^{2it}\partial_t=\partial_\tau,
\eea
so that
\bea
{\widehat \Omega}_{+1}\mapsto \widetilde{ \Omega}_{+1}=e^{S_1}{\widehat \Omega}_{+1}e^{-S_1}= ie^{2it}\partial_t -\widetilde{H}_{+1},
\eea
with
\bea
{\widetilde H}_{+1}&=& e^{2it}\left( i{\widehat X}_++e^{t{\widehat X}_+}{\widehat H}_0e^{-t{\widehat X}_+}\right).
\eea
Due to the commutators
\bea
\relax [{\widehat X}_+, {\widehat H}_0] = 2i {\widehat K}_+, \quad &\quad&\quad [{\widehat X}_+,{\widehat  K}_+]= -2i{\widehat  K}_+,
\eea
where
\bea
{\widehat K}_+&=&\frac{1}{2}(\partial_x+x)^2-i\gamma x\partial_y,
\eea
we obtain
\bea
{\widetilde H}_{+1}&=& e^{2it} \left(i{\widehat X}_++{\widehat H}_0+{\widehat K}_+\right)-{\widehat K}_+.
\eea
The remarkable identity
\bea 
i{\widehat X}_++{\widehat H}_0+{\widehat K}_+&=&0
\eea 
implies that ${\widetilde H}_{+1}$ does not depend on the time coordinate (either $t$ or $\tau$).\par
The second similarity transformation (induced by $S_2$) allows to express
\bea 
\widetilde{\Omega}_{+1}\mapsto {\overline \Omega}_{+1} &=& e^{S_2}\widetilde{\Omega}_{+1}e^{-S_2}= i\partial_\tau+\frac{1}{2}\partial_x^2-i\gamma x\partial_y
\eea
in the form which reduces, in the $\gamma\rightarrow 0$ limit, to the standard free Schr\"odinger equation in $1+1$ dimensions.\par
One should observe that the similarity transformation preserves the symmetry properties of the equations, mapping first-order invariant operators into first-order invariant operators.\par
The following commutative diagram is obtained:

\bea\label{commdiag}
&
\begin{array}{ccc}
~~~~coupled ~(\gamma\neq0):~~ \quad\quad {\bf{Free}}_\gamma ^{0,\pm1}(\tau)&
 \stackrel{{\bf S}}{\longleftrightarrow}&{\bf Osc}_\gamma^{0,\pm 1}(t)\\
 \quad\quad\quad\quad\quad\quad\quad\quad\quad\quad~^{_{{\bf r}}}\downarrow\quad&&\downarrow^{_{{\bf r}}}\\
decoupled ~(\gamma=0):\quad~ ~~~{\bf Free}^{0,\pm 1}~ (\tau)&\stackrel{{\bf S}}{\longleftrightarrow}&{\bf Osc}^{0,\pm 1}(t)
\end{array}&
\eea
The left (right) part of the diagram  denotes the equations obtained from the  differential realizations (\ref{freerep}) and (\ref{oscrep}), respectively.  The horizontal arrows indicate the similarity transformation together with the change of the time coordinate, $\tau$ and $t$ respectively. \par
The three invariant PDEs (at degree $0, \pm 1$) are mapped into each other. \par
In the left part, the deformed Schr\"odinger invariant PDEs correspond to $deg~1$ and possess a continuous spectrum. \par
%zzzrev
In the right part the deformed Schr\"odinger invariant PDEs correspond to $deg~0$. They possess a real, discrete spectrum which, as shown in Section {\bf 7}, coincides with the spectrum of two decoupled harmonic oscillators. We will call this system the {\em  $\ell=\frac{3}{2}$ oscillator}.
 \par
%zzzrev
The vertical arrows denote the mapping to the decoupled systems.  This mapping can be reached in two ways:\par
{\em i}) the singular similarity transformation 
\bea\label{notinv}
&g\mapsto R_1gR_1^{-1},  \quad { \textstyle with} \quad R_1= e^{\alpha y\partial_y}&
\eea 
(such that $\gamma\rightarrow e^{-\alpha}\gamma) $ in the $\alpha\rightarrow\infty $ limit.\footnote{For $\alpha$ real the transformation (\ref{notinv}) is not unitary. The choice $\alpha=\ln \gamma$ allows to set $\gamma=1$. The free parameter $\gamma\in ]0,+\infty[$ labels different unitarily inequivalent theories with testable consequences. See, e.g., the probability decay to the vacuum state presented in (\ref{vacuumdecay}).}\par Despite the singularity of the limit, the invariant equations of the upper part of the diagram admits as non-singular limit the decoupled equations of the lower part of the diagram.  This similarity transformation preserves the symmetry of the equations, mapping first-order invariant operators into first-order invariant operators;\par
{\em ii}) the non-singular similarity transformation
\bea\label{nonsingular}
&g\mapsto R_2gR_2^{-1},  \quad R_2 =e^{\textstyle{({\textstyle\frac{3i}{8}} \gamma x+ {\textstyle{\frac{i}{8}}\gamma\partial_x-{\textstyle{\frac{1}{96}}}\gamma^2
\partial_y)\partial_y}}}.&
\eea
This non-singular transformation does not preserve the symmetry of the equation because some of the transformed generators are no longer first-order differential operators. \par
%Nevertheless, it proves that
%the deformed cryptohermitian operators possess the same spectrum of eigenvalues  as the $\gamma=0$ %decoupled  cryptohermitian operators of the same frequency.\par
The four Schr\"odinger equations associated with the commutative diagram, starting from the upper right corner and proceeding clockwise, are:  (I) the $\ell=\frac{3}{2}$ oscillator (\ref{cryptoosc}),  (II) the decoupled (i.e. $\gamma=0$) $\ell=\frac{3}{2}$ oscillator, (III) the free Schr\"odinger equation in $1+1$ dimensions and, finally, (IV) the deformed free Schr\"odinger equation (\ref{deforfree}). \\
The three inequivalent (with constant, linear and quadratic potential, see \cite{Toppan:2015qla,Niederer:1972zz,Niederer:1973tz,Niederer:1974ba}) Schr\"odinger equations in $1+1$ dimensions invariant under the Schr\"odinger algebra are recovered as restrictions of the $\ell=\frac{3}{2}$ invariant PDEs.  Indeed, if we introduce the $x,y$ separation of variables, the equation of the harmonic oscillator and the free Schr\"odinger equation are recovered by setting $\partial_y\equiv 0$ from, respectively, equations (I)  and (IV). The linear Schr\"odinger equation is recovered from equation (IV)  after setting $\Psi(\tau,x,y) = \psi(\tau,x)\phi(y)$, with the restriction $\partial_y \phi(y)= k\phi(y)$.

\section{The general ${\ell} =\frac{3}{2}$ oscillator}

The ${\widehat{\mathfrak{cga}}}_{\frac{3}{2}}$ conformal Galilei invariance  requires the coupling parameter $\gamma\neq 0$. Since a unitary transformation changes its phase, we can assume
without loss of generality that $\gamma$ belongs to the fundamental domain $\gamma\in ]0,+\infty [$. \par
For $\gamma$ real, the invariant PDEs in the left part of the commutative diagram (\ref{commdiag}) are hermitian. This is not the case for the invariant PDEs in the right part of the diagram. Under hermitian conjugation, the $\ell=\frac{3}{2}$ oscillator equation is transformed into its conjugate
\bea
{\widehat \Omega}_0^\dagger(\gamma)\Psi(t,x,y) =0   &\equiv&( i\partial_t +\frac{1}{2}{\partial_x^2}-\frac{1}{2}x^2 +3 y\partial_y-i\gamma x\partial_y+\frac{3}{2})\Psi(t,x,y).
\eea
All operators
\bea\label{defosc}
\Theta &=& -\frac{1}{2}\partial_x^2 +\frac{1}{2}x^2 +\omega y\partial_y -i\gamma x\partial_y + C,
\eea
for any arbitrary constant $C$ and any $\gamma\neq 0$, induce a Schr\"odinger equation with $\ell=\frac{3}{2}$ Conformal Galilei symmetry, if $\omega$ is restricted to the values 
\bea
\omega&=&\pm \frac{1}{3}, \pm 3.
\eea
The $\omega\leftrightarrow -\omega$ change of sign is explained by the hermitian conjugation. Understanding the $\omega\leftrightarrow
\frac{1}{\omega}$ transformation is subtler.  One should note at first that in the $\gamma=0$ decoupled case the role of the space coordinates $x,y$ can be exchanged by performing the canonical transformation 
\bea
  y \ \leftrightarrow \ \frac{1}{\sqrt{2}} (x-\partial_x),&&\partial_y \ \leftrightarrow \ \frac{1}{\sqrt{2}} (x+\partial_x).  \nonumber
\eea
Next, the coupling term is introduced in terms of the non-singular similarity transformation
given by the inverse of equation (\ref{nonsingular}).  As it turns out, this procedure guarantees
the conformal Galilei invariance of the resulting PDE.\par
An explicit check of the symmetries of this class of PDEs proves that, in order to have the on-shell invariant equations $[{\widehat z}_\pm, {\widehat \Omega}_0]= f_\pm\cdot{\widehat \Omega}_0$, with $f_\pm$ arbitrary functions of the coordinates and symmetry generators of the form
${\widehat z}_\pm =e^{\pm i\lambda t}(\partial_t + {X}_\pm)$, (${X}_\pm$ time-independent operators and $\lambda\neq 0$), the following necessary and sufficient condition has to be satisfied:  the two equations   
\bea
\lambda(\omega^2+1-\frac{5}{2}\lambda^2)&=&0,
\nonumber\\
-3\lambda^2+3\lambda^4+2\lambda\omega+
4\lambda^3\omega-\lambda^2\omega^2-2\lambda\omega^3&=&0,
\eea 
must be simultaneously solved. The only non-vanishing solutions for $\lambda$ are encountered at  $\omega=\pm 3$ and $\omega=\pm\frac{1}{3}$. Therefore, the $\omega=\pm \frac{1}{3}, \pm 3$ critical values are special points of enhanced symmetry. 

\section{Symmetry of the decoupled $\ell=\frac{3}{2}$ oscillator}

By applying the same considerations as in Section {\bf 4}, it is sufficient to analyze the symmetry of the decoupled ($\gamma=0$) $\ell=\frac{3}{2}$ operator
\bea
{\Omega} &=& i\partial_t +\frac{1}{2}\partial_x^2-\frac{1}{2}x^2 +\omega y\partial_y
\eea
in the range $\omega\in [1,\infty [$.\par
For a generic $\omega$ the following invariant operators can be encountered at degree $0, \pm\frac{1}{2}, \pm\frac{\omega}{2},\pm 1$:
\bea\label{generic}
{z}_\pm &=& e^{\pm2i t} (\partial_t\pm ix\partial_x +i\omega y\partial_y+ix^2\pm {\textstyle{\frac{i}{2}}}),\nonumber\\
{ z}_0&=& \partial_t +i\omega y\partial_y,\nonumber\\
{d}&=& -\frac{i}{2}\partial_t,\nonumber\\
{c}&=&1,\nonumber\\
{w}_{\omega}&=& e^{i\omega t} \partial_y,\nonumber\\
{w}_{1}&=& e^{it}(\partial_x +x),\nonumber\\
{w}_{-1}&=& e^{-it}(\partial_x -x),\nonumber\\
{w}_{-\omega}&=& e^{-i\omega t} y.
\eea
${ d}$ is the degree operator. Explicitly, the degree is
\bea
&\pm 1:~ {z}_\pm;\quad
0:~{ z}_0, { d}, { c};\quad
\frac{\pm \omega}{2}: ~{w}_{\pm \omega};\quad
\quad \frac{\pm 1}{2}: ~{w}_{\pm 1}.&
\eea
This $9$-generator symmetry algebra closes the ${\mathfrak{u}}(1){
\supset \hspace{-1em}\hspace{-1pt}+}
(\mathfrak{sch}(1)\oplus{\mathfrak{h}}(1))$ algebra, with non-vanishing commutation relations given by
\bea
\relax [{d}, { z}_\pm ] &=& \pm { z}_\pm,\nonumber\\
\relax [ {d}, {w}_k]&=& \frac{k}{2}{w}_k,\nonumber\\
\relax [{z}_0,{z}_\pm ] &=& \pm 2i { z}_\pm,\nonumber\\
\relax [{z}_+,{ z}_-] &=& - 4 i { z}_0,\nonumber\\
\relax [ {z}_0, {w}_{\pm 1}]&=& \pm i {w}_{\pm 1},\nonumber\\
\relax [{ z}_\pm, {w}_{\mp 1}]&=& \mp 2i {w}_{\pm 1},\nonumber\\
\relax [{w}_1,{w}_{-1}] &=& - 2 {c},\nonumber\\
\relax [w_\omega, w_{-\omega}]&=& {c}.
\eea
$ {d}$ is the generator of the ${\mathfrak{u}}(1)$ subalgebra, while $ {z}_0, {z}_{\pm}, {w}_{\pm 1}, {c} $ generate the Schr\"odinger algebra $ \mathfrak{sch}(1) $ 
and $ w_{\pm \omega}, {c} $ generate the Heisenberg algebra $ {\mathfrak{h}}(1). $ \\
The critical values $\omega=1$ and $\omega =3$ are points of enhanced symmetry for the decoupled system.\par
\subsection{The enhanced symmetry for the decoupled $\omega=1$ system}

At the critical value $\omega=1$ three extra generators are found at degree $0$ and $-1$:
\bea
q_1 &=& y(\partial_x+x),\nonumber\\
q_2 &=& e^{-2it}y^2,\nonumber\\
q_{3}&=& e^{-2it} y(\partial_x-x).
\eea
They have to be added to the previous set of (generic) symmetry generators
\bea
{ z}_\pm &=& e^{\pm2i t} (\partial_t\pm ix\partial_x +i y\partial_y+ix^2\pm {\textstyle{\frac{i}{2}}}),\nonumber\\
{z}_0&=& \partial_t +i y\partial_y,\nonumber\\
{d}&=& -\frac{i}{2}\partial_t,\nonumber\\
{c}&=&1,\nonumber\\
{w}_{1b}&=& e^{i t} \partial_y,\nonumber\\
{w}_{1a}&=& e^{it}(\partial_x +x),\nonumber\\
{w}_{-1a}&=& e^{-it}(\partial_x -x),\nonumber\\
{ w}_{-1b}&=& e^{-i t} y,
\eea
where we denoted, for $\omega =1$, $w_{\pm\omega}$ entering (\ref{generic}) as ``$w_{\pm 1b}$". \par
The extra non-vanishing commutation relations involving the $q_i$'s generators are
\bea
\relax [{ z}_0,q_1]&=& iq_1,\nonumber\\
\relax [ {d}, q_2] &=& - q_2,\nonumber\\
\relax[{d}, q_3] &=& -q_3,\nonumber\\
\relax[{z}_+, q_3]&=& -2i q_3,\nonumber\\
\relax [ {z}_-, q_1]&=& 2i q_3,\nonumber\\
\relax[ { w}_{1b}, q_1]&=& {w}_{1a},\nonumber\\
\relax[{ w}_{-1a}, q_1] &=& 2{ w}_{-1b},\nonumber\\
\relax[ { w}_{1b}, q_2]&=& 2{ w}_{-1b},\nonumber\\
\relax [ {w}_{1b}, q_3] &=& {w}_{-1a},\nonumber\\
\relax [ {w}_{1a}, q_3] &=& -2 { w}_{-1b},\nonumber\\
\relax [ q_1, q_3] &=& -2q_2.
\eea
The symmetry algebra closes as a non semi-simple, $12$-generator, Lie algebra.

\subsection{The enhanced symmetry for the decoupled $\omega=3$ system}

At the $\omega =3$ critical value the three extra generators
${ r}_{-j}$, $j=1,2,3$, of degree $-j$, are encountered.
We have, explicitly,
\bea\label{rgenerators}
{r}_{-1} &=& e^{-2it} y(\partial_x+x),\nonumber\\
{ r}_{-2} &=& e^{-4 it} y(\partial_x -x),\nonumber\\
{ r}_{-3} &=& e^{-6 i t} y^2.
\eea
At $\omega=3$ the symmetry algebra is a 12-generator algebra which differs from the $12$-generator symmetry algebra of the $\omega=1$ decoupled system.\par
 The extra non-vanishing commutation relations involving the ${r}_{-j}$ generators are given by
\bea
\relax [{d}, {r}_{-j}] &=& -j { r}_{-j},\nonumber\\
\relax [{z}_0, { r}_{-1}] &=& i { r}_{-1},\nonumber\\
\relax [{z}_0, { r}_{-2}]& =& -i {r}_{-2},\nonumber\\
\relax [{ z}_{-}, {r}_{-1}]& =& 2i {r}_{-2},\nonumber\\
\relax [{z}_{+}, {r}_{-2}]& =& -2i {r}_{-1},\nonumber\\
\relax [{ w}_{+3}, {r}_{-1}]& =& {w}_{+1},\nonumber\\
\relax [{w}_{-1}, {r}_{-1}]& =& 2 { w}_{-3},\nonumber\\
\relax [{w}_{+3}, {r}_{-2}]& =& { w}_{-1},\nonumber\\
\relax [{w}_{+1}, {r}_{-2}]& =& -2 {w}_{-3},\nonumber\\
\relax [{w}_{+3}, { r}_{-3}]& =& 2 {w}_{-3},\nonumber\\
\relax [{r}_{-1}, { r}_{-2}]& =& -2 {r}_{-3}.
\eea

\section{The contraction algebra}

A contraction algebra is recovered from (\ref{oscrep}) 
by taking the $\gamma\rightarrow 0$ limit and by suitably rescaling the generators. The contraction requires the rescaling ${\widehat g}\mapsto {\widetilde g}= \gamma^s {\widehat g}$
(${\widehat g}$ is any generator entering (\ref{oscrep})), with the power $s$ given by
\bea
s=0 &:& {\widehat z}_0, {\widehat z}_+, {\widehat w}_3, {\widehat c},\nonumber\\
s=1 &:& {\widehat z}_-, {\widehat w}_1, {\widehat w}_{-1},\nonumber\\
s=2&:& {\widehat  w}_{-3}.
\eea
The contracted $8$-generator algebra expressed by ${\widetilde z}_\pm, {\widetilde z}_0, {\widetilde c}, {\widetilde w}_k$ ($k=\pm1, \pm 3$) is a subalgebra of the full $12$-generator symmetry algebra of the $\omega=3$ decoupled system with generators given by (\ref{generic}) and (\ref{rgenerators}). The identification goes as follows
\bea
{\widetilde z}_+ &=& e^{\widetilde S}{z}_+e^{-{\widetilde S}}= e^{2it}(\partial_t+ ix\partial_x +3iy\partial_y+ix^2+2i),\nonumber\\
{\widetilde z}_0 &=& e^{\widetilde S}(2i{d}-{\textstyle{\frac{3i}{2}}}{c})e^{-{\widetilde S}}= \partial_t,\nonumber\\
{\widetilde z}_- &=& e^{\widetilde S} (12 i{ r}_{-1})e^{-{\widetilde S}}= 12 i e^{-2it} y(\partial_x+x),\nonumber\\
{\widetilde w}_{+3} &=& e^{\widetilde S}{w}_{+3}e^{-{\widetilde S}}= e^{3it}\partial_y,\nonumber\\
{\widetilde w}_{+1} &=& e^{\widetilde S} (-2i{ w}_{+1})e^{-{\widetilde S}}= 2i e^{it} (\partial_x+x),\nonumber\\
{\widetilde w}_{-1} &=& e^{\widetilde S} (-4i{w}_{-1})e^{-{\widetilde S}}= 4i e^{-it} (\partial_x-x),\nonumber\\
{\widetilde w}_{-3} &=& e^{\widetilde S} (48{w}_{-3})e^{-{\widetilde S}}= -48 e^{-3it} y,\nonumber\\
{\widetilde c} &=& e^{\widetilde S} {c}e^{-{\widetilde S}}= 1,
\eea
with the similarity transformation  given by ${\widetilde S}= -\textstyle{\frac{3}{2}}it$.\par
The contraction algebra corresponds to the two-dimensional Euclidean algebra acting on {{two sets}} of creation/annihilation operators.
We have the
${\mathfrak{e}}(2){\supset \hspace{-1em}\hspace{-1pt}+}{\mathfrak{h}}(2)$ algebra, with non-vanishing commutators given by
\bea
\relax [{\widetilde z}_0,{\widetilde z}_{\pm}]&=& \pm 2i{\widetilde z}_\pm,\nonumber\\
\relax[ {\widetilde z}_0, {\widetilde w}_{k}]&=& k{\widetilde w}_k,\nonumber\\
\relax[{\widetilde z}_+,{\widetilde w}_{-1}]&=& -4i {\widetilde w}_{+1},\nonumber\\
\relax [ {\widetilde z}_-, {\widetilde w}_{+3}]&=& 6i {\widetilde w}_{+1},\nonumber\\
\relax[{\widetilde z}_-,{\widetilde w}_{-1}] &=& 2i {\widetilde w}_{-3},\nonumber\\
\relax[{\widetilde w}_{|k|}, {\widetilde w}_{-|k|}]&=& (3-2|k|) 16{\widetilde c}.
\eea

\section{Non-hermitian deformed oscillators with real eigenvalues}

The non-hermitian operator derived from the degree $0$ invariant equation (\ref{cryptoosc}) is
\bea\label{cryptoosc2}
H_0(\gamma)&=&
-\frac{1}{2}{\partial_x}^2+\frac{1}{2}x^2 +3y\partial_y+i\gamma x\partial_y+\frac{3}{2}.
\eea
It may be written (see (\ref{oscrep})) in the form
\begin{equation} \label{cryptHamil1}
  H_0(\gamma) = \frac{\gamma^2}{16} ({\widehat w}_{-1}{\widehat  w}_{+1} -{\widehat  w}_{-3}{\widehat  w}_{+3}) + 2.
\end{equation}
{{A discrete spectrum can be calculated algebraically from a lowest weight representation.}}
The ground state $\psi_{0,0} $ is defined by
\begin{equation} \label{Def:GS}
   {\widehat w}_{+1} \psi_{0,0} = {\widehat w}_{+3} \psi_{0,0} = 0.
\end{equation}
The solution of the above equations gives the explicit expression of the ground state (up to normalization)
\bea
 \psi_{0,0} &=& e^{-x^2/2}. 
\eea  
The excited states are given by
\begin{equation} \label{Def:ExS}
  \psi_{n,m} = {\widehat w}_{-1}^n {\widehat w}_{-3}^m \psi_{0,0}.
\end{equation}
The corresponding eigenvalue $E_{n,m}$ is computed by the commutation relations
\begin{equation} \label{SectGenAlg}
  [H_0(\gamma), {\widehat w}_{-1}] = {\widehat w}_{-1}, \qquad 
  [H_0(\gamma),{\widehat w}_{-3}] = 3{\widehat w}_{-3}.
\end{equation}
One finds that the operator $ H_0(\gamma) $ has real discrete eigenvalues which are identical to the decoupled harmonic oscillator eigenvalues:
\bea 
E_{n,m}&=& n+3m +2. \label{Eigen}
\eea
The explicit form of the eigenfunctions are readily obtained from (\ref{oscrep}). We list some of them.
\bea
   \psi_{1,0} &=& {\widehat w}_{-1} \psi_{0,0} = -\frac{8i}{\gamma} x e^{-x^2/2-it},
   \nonumber\\
   \psi_{2,0} &=& {\widehat w}_{-1}^2 \psi_{0,0} =  2\Big( \frac{4i}{\gamma} \Big)^2 (2x^2-1) e^{-x^2/2-2it},
  \nonumber \\
   \psi_{0,1} &=& {\widehat w}_{-3} \psi_{0,0} = -\frac{24}{\gamma^2} (i\gamma x+ 2y) e^{-x^2/2-3it},
  \nonumber \\
   \psi_{1,1} &=& {\widehat w}_{-1} {\widehat w}_{-3} \psi_{0,0} = \frac{48}{\gamma^2} \Big( 1 -4x^2 + \frac{4i}{\gamma}xy \Big) e^{-x^2/2-4it}.
\eea
In general, the excited state wavefunctions have the form of
\begin{equation} \label{ExStWaveF}
  \psi_{n,m} = {\cal P}_{n,m}(x,y) e^{-x^2/2} e^{-(n+3m)it},
\end{equation}
where $ {\cal P}_{n,m}(x,y) $ is a degree $n+m$ polynomial in $x, y. $ 
The phase factor $ e^{-(n+3m)it} $ is not essential to the eigenvalue problem. \par
Following 
\cite{smilga2009comments}, we can call ``cryptohermitian" any non-hermitian operator (such as (\ref{cryptoosc2})) with a real spectrum. The operator (\ref{cryptoosc2}), being invariant under the transformations $x\mapsto -x, y\mapsto y, i\mapsto -i$, is PT-symmetric.  Therefore, it belongs to the class of non-hermitian PT-symmetric operators (see \cite{beboe} and \cite{beboe2} and \cite{ben} for a review) with real spectrum. It was pointed out in
\cite{mos1}, 
\cite{mos2} and
\cite{mos3} that PT-symmetric operators are often pseudo-hermitian. It turns out that this is not the case for the operator (\ref{cryptoosc2}), since the similarity transformation (\ref{nonsingular}) is not realized by a hermitian operator. For this reason the stronger notion of quasi-hermiticity, as well as the constructions based upon that (see \cite{sgh}, \cite{mobat} and \cite{bazno}),  is not applicable in our context. \par
In Section {\bf 8} we discuss the connection of the operator (\ref{cryptoosc2}) with a special class of Pais-Uhlenbeck oscillators, recovered at given algebraic frequencies. It is worth mentioning that PT-symmetry in the context of Pais-Uhlenbeck oscillators (at arbitrary, unequal, frequencies) was investigated in \cite{beman}, 
\cite{mann} and \cite{limi}.\par
In order to pave the way to discuss the connection between Conformal Galilei Algebras and Pais-Uhlenbeck
oscillators at special frequencies, we need to introduce the operator $K({\overline\gamma})$, obtained from
(\ref{cryptoosc2}) via a non-canonical transformation which preserves the canonical commutation relations, but does not preserve the hermitian conjugation property. Up to a vacuum energy constant, we can set 
\bea\label{herosc}
K({\overline\gamma})&=& a^\dagger a+3 b^\dagger b +\frac{1}{2}+{\overline \gamma}(a+a^\dagger)b,
\eea
through the positions
\bea
&a=\frac{1}{\sqrt{2}}(x+\partial_x),\quad a^\dagger =\frac{1}{\sqrt{2}}(x-\partial_x),\quad b=
\frac{1}{\sqrt{2}}(z+\partial_z),\quad b^\dagger =
\frac{1}{\sqrt{2}}(z-\partial_z),\quad {\overline\gamma}=\frac{i}{\sqrt{2}}\gamma.&
\eea
The
two independent creation/annihilation operators ($a, a^\dagger$ and $b, b^\dagger$) have non vanishing commutators: $[a,a^\dagger]=[b,b^\dagger]=1$. \par
The similarity transformation (\ref{nonsingular}) maps $ K({\overline\gamma}) $ into the decoupled hermitian operator $K(0)$ through
\bea\label{Kgamma}
K({\overline\gamma}) &=& R K(0) R^{-1},  \quad  R =e^{-( {\frac{{\overline\gamma}}{2}}a^\dagger b+\frac{\overline\gamma}{4}ab+\frac{{\overline\gamma}^2}{48}b^2)}.
\eea
Since, as mentioned before, $R$ is not hermitian, $K({\overline{\gamma}})$, for ${\overline\gamma}\neq 0$, is not a pseudo-hermitian operator.\par
% We can identify $b\equiv y$, with $y$ the coordinate entering (\ref{cryptoosc2}).\par
The operator $K({\overline\gamma})$ acts on the {{Hilbert space ${\mathcal L}^2({ {\mathbb R}^2})$}}, the space of square-integrable functions defined on the coordinates $x,z$. \par
%The operators  $H_0(\gamma)$ and $K(\overline \gamma)$ (which act on different Hilbert spaces) are,
%up to a vacuum energy constant, isospectral.
%zzzrev
% The  $K({\overline \gamma})$ eigenvectors belong to ${\mathcal L}^2({\widetilde {\mathbb R}^2})$. \par \
% They are given by the (unnormalized) states $|n,m> = (a^\dagger)^n(b^\dagger)^m|vac>$, where 
% $| vac>\equiv |0,0>$ is the Fock vacuum defined by the conditions $a|vac>=b|vac>=0$.
\par
One can read from the commutators
\bea\label{mode}
[K(\overline\gamma), A_\lambda] &=& \lambda A_\lambda
\eea
which excited modes are created.\par
For any ${\overline \gamma}\neq 0$, the solutions of the (\ref{mode}) equation are obtained for $\lambda=\pm 3, \pm\frac{1}{3}$. The corresponding modes are
\bea\label{bogoljubovlike}
A_{-3}&=& b,\nonumber\\
A_{-1}&=& a+\frac{1}{2}{\overline\gamma}b,\nonumber\\
A_{+1}&=& a^\dagger-\frac{1}{4}{\overline\gamma}b,\nonumber\\
A_{+3}&=& b^\dagger-\frac{1}{2}\overline\gamma a^\dagger -\frac{1}{4}\overline\gamma a+\frac{1}{24}\overline\gamma^2b.
\eea
In this basis the non-vanishing commutators are
\bea
\relax [A_{-i},A_j]&=&\delta_{ij}, \quad\quad\quad (i,j=1,3).
\eea
The non-hermitian operator $K(\overline\gamma)$ commutes with the ``non-hermitian analog of the Number operator", $N(\overline\gamma)$. In terms of the $A_k$ modes, the operators are given by
\bea
K(\overline\gamma)&=& 3A_3A_{-3}+A_1A_{-1}+\frac{1}{2},\nonumber\\
N(\overline\gamma)&=& A_3A_{-3}+A_1A_{-1},\nonumber\\
\relax [K(\overline\gamma),N(\overline\gamma)]&=&0.
\eea
The Fock vacuum $|vac>$ satisfies
\bea
a|vac>=b|vac>=0&,& A_{-1}|vac>=A_{-3}|vac>=0.
\eea
The Hilbert space ${\mathcal L}^2({{\mathbb R}^2})$ can be spanned by both sets of (unnormalized) states,
\bea
|n,m> &=& (a^\dagger)^n(b^\dagger)^m|vac>,\nonumber\\
|\overline n,\overline m>&=& A_1^nA_3^m|vac>.
\eea
The invertibility of the Bogoliubov-type transformation (\ref{bogoljubovlike}) implies that the
$|\overline n,\overline m>$ eigenstates form a complete (albeit non-orthogonal) set which can be expressed in terms of the decoupled eigenvectors $|n,m>$ (the converse is also true).\par
We can therefore write
\bea
&|vac>=|0,0>=|\overline 0,\overline 0>.&
\eea
The spectrum of $K(\overline\gamma)$, $N(\overline\gamma)$ coincides with the spectrum of the Hamiltonian and Number operator of a decoupled harmonic oscillator.
$|\overline n,\overline m> $ is an eigenvector for $K(\overline\gamma)$, $N(\overline\gamma)$ with respective eigenvalues $n+3m+\frac{1}{2}$ and $n+m$. In increasing order of $K(\overline\gamma)$ eigenvalues, the first (unnormalized) common eigenvectors of $K(\overline\gamma)$, $N(\overline\gamma)$ are
\bea
(\frac{1}{2},0): && |\overline 0,\overline 0>=|0,0>=|vac>,\nonumber\\
(\frac{3}{2},1): && |\overline 1,\overline 0>=|1,0>,\nonumber\\
(\frac{5}{2},2): && |\overline 2,\overline 0>=|2,0>,\nonumber\\
(\frac{7}{2},1): && |\overline 0,\overline 1>=|0,1>-\frac{1}{2}\overline\gamma|1,0>,\nonumber\\
(\frac{7}{2},3): && |\overline 3,\overline 0>=|3,0>,\nonumber\\
(\frac{9}{2},2): && |\overline 1,\overline 1>=|1,1>-\frac{1}{2}\overline\gamma |2,0>-\frac{1}{4}\overline\gamma|0,0>,\nonumber\\
(\frac{9}{2},4): && |\overline 4,\overline 0>=|4,0>.
\eea
Since the operators are non-hermitian, their eigenvectors are non-orthogonal. This implies measurable physical consequences.
Let us suppose  that we are able to prepare the system in a given common eigenvector of $K(\overline\gamma)$, $N(\overline\gamma)$, let's say the state $|\overline 1,\overline 1>$. Following the standard rule of Quantum Mechanics we can compute the probability for this state to collapse, after a measurement operation, to the vacuum state. A simple computation shows that the probability is $p=|_N<\overline 1,\overline 1|\overline0,\overline0>_N|^2 $ ($|\overline 0,\overline 0>_N$, $|\overline 1,\overline 1>_N$ are the normalized states). In the case at hand we have 
\bea\label{vacuumdecay}
p&=&\frac{|\overline \gamma|^2}{16+9|\overline\gamma|^2}.
\eea
This probability is restricted in the range $0\leq p<\frac{1}{9}<1$.
The deformation coupling parameter ${\overline \gamma}$, via its squared modulus,  has testable consequences. 

\section{Comment on  Pais-Uhlenbeck oscillators and the $\ell\geq{\textstyle{\frac{5}{2}}} $ cases}

The same spectrum of eigenvalues is obtained for\\
{\em i}) the coupled ($\gamma\neq 0$) cryptohermitian operator (\ref{cryptoosc2}), \\
{\em ii}) the decoupled ($\gamma=0$) cryptohermitian operator and  (up to a vacuum energy shift) \\
{\em iii}) the hermitian
Hamiltonian (given by (\ref{herosc}) for ${\overline\gamma}=0$) of two decoupled oscillators.\par
The construction of Section {\bf 7} can be repeated by starting with the hermitian conjugate of the (\ref{cryptoosc2}) operator. In this case the spectrum of the three resulting operators is unbounded. It is given, up to the vacuum energy shift, by $E_{n,m}= n-3m$. The Hilbert space of the decoupled harmonic oscillators with energy modes $1,-3$ continues to be {{ ${\mathcal L}^2({ {\mathbb R}^2}$)}}, obtained by applying the creation operators $a^\dagger, b^\dagger$ to the Fock state $|0,0>$ ($a|0,0>=b|0,0>=0$).
Due to the unboundedness of the spectrum, $|0,0>$ can no longer be interpreted as the vacuum state.\par
The system with unbounded spectrum is related to the Pais-Uhlenbeck oscillators. We recall \cite{pais1950field,smilga2009comments} that the Pais-Uhlenbeck model is a higher derivative system. It admits, via the Ostrogradski\u{\i} construction \cite{ostrogradski1850member} (see \cite{Woodard:2015zca} for a review), 
a Hamiltonian formulation. The resulting Ostrogradski\u{\i} Hamiltonian is canonically equivalent to a set of decoupled harmonic oscillators with alternating (positive and negative) energy modes. The $n$-oscillator Pais-Uhlenbeck system is canonically expressed as
\bea
H_n &=& \sum_{i=1}^{n} (-1)^{i+1} \omega_i a_i^\dagger a_i,
\eea
where $\omega_i\in {\mathbb R}$ and the constraint $\omega_i<\omega_{i+1}$ is satisfied.\par
The harmonic oscillator with energy modes $1,-3$ is a special case of the $2$-oscillator Pais-Uhlenbeck model. 
In a series of papers \cite{galajinsky2013dynamical2,andrzejewski2014conformal,andrzejewski2014conformal2,andrzejewski2014hamiltonian,galajinsky2015dynamical} 
the Pais-Uhlenbeck oscillators with energy modes given (up to a normalization factor) by the arithmetic progression $\omega_i = 2i-1$ were linked to the Conformal Galilei Algebras ${\widehat{\mathfrak{cga}}}_\ell$ (with $\ell = n-\textstyle{\frac{1}{2}}$).\par
The present analysis proves that this association is rather subtle.
%zzzrev
We would like to stress that in this paper we consider (as it is standard in PDE's theory) the symmetry generators to be at most {\em first-order} differential operators. In this respect the Pais-Uhlenbeck PDE given by the decoupled harmonic oscillators does not possess any enhanced symmetry (not even at the special $1,-3$ energy modes).  The PDE, invariant under the Conformal Galilei Algebra, is obtained for the coupled operator only for $\gamma\neq 0$. The decoupled operator $(\gamma=0)$
possesses the 12-generator symmetry algebra (introduced in Section {\bf 6}) which does not contain the Conformal Galilei Algebra as a subalgebra. \par
Even so, these results are not in contradiction with the findings in previous works on Pais-Uhlenbeck oscillators.  For example, 
in \cite{andrzejewski2014conformal} it was shown that on the Hamiltonian level the $\omega_1=1,\omega_2=-3$ Pais-Uhlenbeck oscillators possess the (centrally extended) $\frac{3}{2}$-conformal Galilei symmetry in terms of Poisson brackets among conserved charges. It can be shown, on the other hand, that at the quantum level, some of the Noether charges entering formula (39) in \cite{andrzejewski2014conformal} are {\em second-order} differential operators. \par
%Another paper \cite{galajinsky2015dynamical}, relating the (\ref{cryptoosc}) equation (for $\gamma=2i$) %with the Pais-Uhlenbeck oscillators does not address the question of the first-order differential operators.\par
%In 
%The Conformal Galilei Algebra derivation of the Pais-Uhlenbeck oscillators for the energy modes $1,-3$ %requires two non-trivial passages:
%zzzrev
%\\
%1) to perform the $\gamma\rightarrow 0$ decoupling limit. The decoupled PDE no longer possesses the %Conformal Galilei Algebra as invariance. Its symmetry algebra has been discussed in Section {\bf 6}. The %contraction of the Conformal Galilei Algebra is a symmetry subalgebra;\\
%2) to change the conjugation properties, by replacing the decoupled cryptohermitian operator with the %hermitian decoupled harmonic oscillator.\par%
For general half-integer ${\ell}$, the invariant PDEs which possess the Conformal Galilei algebra ${\widehat{\mathfrak{cga}}}_\ell$ (for a definition, see \cite{negro1997nonrelativistic}) as a symmetry algebra, depend on ${\ell }+{\textstyle{\frac{3}{2}}}$ coordinates. The invariant PDEs are deformations of decoupled equations, depending
on ${\ell}-{\textstyle{\frac{1}{2}}}$ deformation parameters $\gamma_j\neq 0$ ($j=1,\ldots, {\ell}-{\textstyle{\frac{1}{2}}}$). The decoupled systems are recovered in the limit, for any $j$,$\gamma_j\rightarrow 0$.\par
The invariant PDEs with continuous spectrum are
\bea
i\partial_\tau \Psi(\tau, {\vec x})&=&\left(-{\textstyle{\frac{1}{2}}\partial_{x_1}^2+ i\sum_{j=1}^{\ell-{\textstyle{\frac{1}{2}}}}\gamma_jx_j\partial_{x_{j+1}}}\right)\Psi(\tau,\vec{ x}).\label{genfree}
\eea

The invariant PDEs with discrete spectrum are
\bea\label{genosc}
i\partial_t\Psi(t, {\vec x})&=&\left(-{\textstyle{\frac{1}{2}}\partial_{x_1}^2+
{\textstyle {\frac{1}{2}}}x_1^2+\sum_{i=2}^{\ell+{\textstyle{\frac{1}{2}}}}\omega_ix_i\partial_{x_i}+ i\sum_{j=1}^{\ell-{\textstyle{\frac{1}{2}}}}\gamma_jx_j\partial_{x_{j+1}}}\right)\Psi(t,\vec{ x}).
\eea
The energy modes $\omega_i$ are normalized so that $|\omega_i| = 2i-1$. The solution
$\omega_i = \epsilon_i |\omega _i|$ with all positive signs ($\forall i$, $\epsilon_i=+1$) corresponds to the bounded discrete spectrum discussed in \cite{Aizawa:2014uma}. By taking the hermitian conjugation, the solution with
flipped signs, $\epsilon_i =-1$ for all $i$, also leads to a ${\widehat{\mathfrak{cga}}}_\ell$-invariant PDE.\par
An explicit computation of the on-shell condition for $\ell={\textstyle{\frac{5}{2}}}$ (similar to the one presented in Section {\bf 4}), proves that  the ${\widehat{\mathfrak{cga}}}_{\frac{5}{2}}$ invariance is guaranteed by both choices of signs, $\epsilon_2=\pm 1$ and $\epsilon_3=\pm 1$. As explained above, the alternating choice ($\epsilon_2=-1, \epsilon_3=+1$) is related to a special case of three Pais Uhlenbeck oscillators (with $1,-3,5$ energy modes). \\
An open problem is finding a general proof, valid for all half-integer ${\ell}$, that every choice of $\epsilon
_i=\pm 1$ signs lead to the ${\widehat{\mathfrak{cga}}}_l$ symmetry algebra of the PDE equation (\ref{genosc}).
\section{Conclusions}

\par

We summarize, in the Conclusions, the list of new results obtained in this paper. \par
We constructed the most general class of second-order PDEs, invariant under the $d=1$ centrally extended Conformal Galilei Algebras ${\widehat{\mathfrak{cga}}}_\ell$ with half-integer $\ell$, proving that they are Schr\"odinger equations which are deformations of decoupled equations. For $\ell=\frac{3}{2}$ the unique deformation parameter is $\gamma\neq 0$ (the decoupled systems being recovered in the $\gamma\rightarrow 0$ limit).\par
At $\ell=\frac{3}{2}$ the invariant PDEs with discrete spectrum, besides $\gamma$, depend on two frequencies $\omega_1, \omega_2$ (entering the equation in non-symmetric form).  The invariance under ${\widehat{\mathfrak{cga}}}_{\ell=\frac{3}{2}}$ is only recovered at special critical values of the ratio
$r=\frac{\omega_2}{\omega_1}$ given by $r=\pm\frac{1}{3}, \pm 3$. The $r=3$ value reproduces the bounded spectrum presented in \cite{Aizawa:2014uma}, while the negative values $r<0$ produce an unbounded spectrum which coincides with the spectrum of two Pais-Uhlenbeck oscillators at the given ratio $r$. \par
We further investigated the symmetry algebra of the $\gamma=0$ decoupled systems for a generic value of  $r=\frac{\omega_2}{\omega_1}$, obtaining the following results.  Enhanced symmetries are encountered at the critical values $r=\pm\frac{1}{3},\pm 1, \pm 3$. Two inequivalent $12$-generator symmetry algebras are recovered at $r=\pm\frac{1}{3}, \pm 3$ and $r=\pm 1$, respectively. \par
The ${\widehat{\mathfrak{cga}}}_{\ell=\frac{3}{2}}$ Conformal Galilei Algebra  is not a subalgebra of the $12$-generator, $\gamma=0$ and $r=\pm\frac{1}{3},\pm 3$, decoupled symmetry. From ${\widehat{\mathfrak{cga}}}_{\ell=\frac{3}{2}}$, in the $\gamma\rightarrow 0$ limit, a contraction algebra is recovered.
The contraction algebra, see Section {\bf 6}, is an $8$-generator subalgebra of the full  $r=\pm\frac{1}{3}, \pm 3$ symmetry algebra of the $\gamma=0$ decoupled system. \par
As a corollary of this analysis we showed that the contraction algebra, rather than the Conformal Galilei algebra itself, is an enhanced symmetry of the
$r=-3$ decoupled Pais-Uhlenbeck oscillators.\par
Concerning the $\ell=\frac{3}{2}$ PDE with discrete spectrum, we showed that, besides the $r=3$ real bounded spectrum, a real unbounded spectrum
is obtained at $r=-\frac{1}{3}, -3$. For all admissible values $r$ the induced operators on the r.h.s. are not hermitian. They are, nevertheless, PT-symmetric
\cite{beboe}.
 They are not, on the other hand, pseudo-hermitian \cite{mos1}, since the non-singular similarity transformations (see equations (\ref{nonsingular}) and (\ref{Kgamma})), mapping the coupled into the decoupled system, are not realized by a hermitian operator. The non-hermitian operator (\ref{herosc}) acts on
the Hilbert space ${\mathcal L}^2({ {\mathbb R}^2})$. Its real discrete spectrum coincides with the spectrum of two decoupled harmonic oscillators. The Bogoljubov transformations relating the coupled and the decoupled systems are explicitly given.\par
For generic half-integer $\ell$, the number of non-vanishing deformation parameters $\gamma_j$ is $\ell-{\textstyle{\frac{1}{2}}}$. For $\ell=\frac{5}{2}$
the arithmetic progression of $\omega_i$ frequencies entering eq. (\ref{genosc}) and producing the ${\widehat{\mathfrak{cga}}}_{\ell=\frac{5}{2}}$ symmetry
algebra is given by $\pm 1, \pm 3, \pm 5$. The $1,3,5$ sequence corresponds to the bounded spectrum, while the $1,-3,5$ sequence corresponds to a
special case of three Pais-Uhlenbeck oscillators.
\par
The extension of this construction to the $\ell={\textstyle{\frac{1}{2}}}+{\mathbb N}_0$ centrally extended Conformal Galilei Algebras  with $d>1$ (see \cite{negro1997nonrelativistic}) is immediate. The invariant PDEs with discrete spectrum correspond to non-hermitian operators whose spectrum is real and given by $d$ copies of the energy modes created in the $d=1$ case.\par
In the class of systems here investigated, the oscillators turn out to have all different frequencies. Equal frequency oscillators are obtained from the different class of second-order PDEs invariant under the $d=2$ centrally extended Conformal Galilei Algebras 
with integer $\ell$,
{{see \cite{aikutoexotic}.}}
\\ {~}~{~}
\par {\Large{\bf Acknowledgements}}
{}~\par{}~\par 
We thank T. Yumibayashi for helpful discussions on Pais-Uhlenbeck oscillators.
Z.K. and F.T. are grateful to the Osaka Prefecture University, where this work was elaborated, for hospitality.
F.T. received support from CNPq (PQ Grant No. 306333/2013-9). 
N. A. is supported by the  grants-in-aid from JSPS (Contract No. 26400209).
%
%
%\begin{thebibliography}{99}
%zzzz
%\end{thebibliography}
\bibliographystyle{aip}
\bibliography{CGA}

\end{document}